\documentclass[preprint, showpacs, preprintnumbers,amsmath,amssymb,nofootinbib]{revtex4}
\usepackage{mathrsfs}
\usepackage{amsmath}
\usepackage[colorlinks, linkcolor=blue, citecolor=red, urlcolor=red]{hyperref}
\usepackage{graphicx}
\usepackage{bm}
\usepackage{psfrag}
\usepackage{amsmath}
\usepackage{amssymb}
\usepackage{latexsym}
\usepackage{exscale}

\newcommand{\modul}[1]{|#1|}

\def\bbm[#1]{\mbox{\boldmath $#1$}}
\newcommand{\bea}[1]{\begin{eqnarray}\label{#1}}
 \newcommand{\eea}{\end{eqnarray}}
 \def\gsim{ \lower .75ex \hbox{$\sim$} \llap{\raise .27ex \hbox{$>$}} }
 \def\lsim{ \lower .75ex \hbox{$\sim$} \llap{\raise .27ex \hbox{$<$}} }
\def\/{\over}
\newcommand{\vect}[1]{\bm{#1}}
\newcommand{\ten}[1]{\mbox{\textbf{
{\textsf{#1}}}}}

\newcommand{\sprod}{\!\cdot\!}

\newcommand{\trans}{\mathsf{T}}
\newcommand{\dif}{\mathrm{d}}
\newcommand{\mi}{\mathrm{i}}

\begin{document}

\title{\bf  The thermalization of a two-level atom  in a planar dielectric system out of thermal equilibrium  }
\author{ Puxun Wu$^{1,2,3}$ and Hongwei Yu$^{1,2}$}
\address{ $^1$Center for Nonlinear Science and Department of Physics, Ningbo
University,  Ningbo, Zhejiang 315211, China\\
$^{2}$Synergetic Innovation Center for Quantum Effects and Applications, Hunan Normal University, Changsha, Hunan 410081, China\\
$^3$Center for High Energy Physics, Peking University, Beijing 100080, China}

\begin{abstract}
We study the thermalization of an elementary quantum system modeled by a two-level atom interacting with stationary electromagnetic fields out of thermal equilibrium near a  dielectric slab. The slab is held at a temperature different from that of the region where the atom is located.  We find that when the slab is a  nonabsorbing and nondispersive dielectric  of a  finite thickness $ d$,  no out of thermal equilibrium effects appear as far as the thermalization of the atom is concerned,
and  a finite thick dielectric slab with a tiny imaginary part in the relative permittivity $\operatorname{Im} \epsilon$ behaves like  a half space dielectric substrate  if $\frac{\operatorname{Im} \epsilon}{\sqrt{\operatorname{Re}\epsilon-1}} \frac{d}{\lambda_0} > 1$ is satisfied, where $\lambda_0$ is the transition wavelength of the atom. This condition can serve as a guide for an experimental verification, using a dielectric substrate of a finite thickness, of the effects that arise from out of thermal equilibrium fluctuations with a half-space (infinite thickness) dielectric.

\end{abstract}


\maketitle
\section{Introduction} 

Physical systems out of thermal equilibrium but in a stationary configuration, such as that of a substrate and an environment  held  respectively at different temperatures,  may exhibit remarkable and measurable quantum phenomena, and thus have recently  attracted  an increasing deal of  interest  both theoretically and experimentally. In this respect,  Antezza et al.~\cite{Antezza}  investigated, in the large distance limit, the Casimir-Polder (CP) force~\cite{Casimir} in such  an out of thermal equilibrium situation. They found  that the CP force shows  new qualitative and quantitative behaviors. Specifically,  the force decays like $1/z^3$  and is proportional to $\Delta T^2$,  where  $z$ is the distance  between an atom and the surface of the substrate, and $\Delta T^2 \equiv T_s^2- T_e^2$ with $T_e$ and $T_s$ being respectively the temperatures of the thermal bath in the right and the substrate in the left half space. This behavior of the force differs clearly from that of an atom both in vacuum which has a $1/z^5$ dependence~\cite{Casimir} and in  a thermal equilibrium environment which behaves like  $T/z^4$ and is attractive~\cite{Lifshitz}.     Actually the out of thermal equilibrium CP force can be either  attractive or repulsive depending on the difference of two temperatures.  Later,  Zhou and Yu analyzed  in detail the behaviors of the out of thermal equilibrium CP force of an atom  near the surface of a half space real dielectric substrate  in different distance regimes~\cite{Zhou014},  where a real dielectric  refers to a nonabsorbing and nondispersive dielectric whose permittivity is real and frequency independent. In addition, the CP force of  a diamagnetic atom  out of thermal equilibrium  has also been investigated in~\cite{Wu14}.   Remarkably, the new behavior of the CP force out of thermal equilibrium has been measured in experiment by  positioning a nearly pure $^{87}$Rb Bose-Einstein condensate a few microns from a dielectric substrate, which   consists of uv-grade fused silica with a $2$~mm thickness~\cite{Obrecht}. 
 
 On the other hand, the dynamics of an elementary quantum system in a stationary environment out of thermal equilibrium, has been studied by Bellomo et al.~\cite{Bellomo}  and it is found that the  quantum system modeled by a two-level atom can be thermalized to a steady state with an effective temperature between the temperature of the wall and that of the environment. The similar result has also been obtained for an atom placed outside a radiating Schwarzschild black hole~\cite{Hu13}. For two quantum emitters interacting with a  common stationary electromagnetic field out of thermal equilibrium,    Bellomo and Antezza found that the absence of equilibrium allows the generation of steady entangled states between the emitters, which is inaccessible at thermal equilibrium~\cite{Bellomo13a,Bellomo13b}. In addition,   the photon heat tunneling was discussed in~\cite{Joulain, Messina12}. Other aspects about the out of thermal equilibrium effects have been discussed in~\cite{Pitaevskii, Antezza08, Buhmann, Sherkunov, Bimonte, Rodriguez, Messina, Kruger, Kruger11,  Noto, Druzhinina, Behunin, Leggio, Zhou15, Antezza06}. 

To simplify the theoretical calculations, a half-space,  even real, dielectric substrate is usually assumed when analyzing the non-equilibrium thermal system.  However, in reality, such a dielectric substrate never exists. In fact,  in experiment, a   dielectric slab with a finite thickness and absorption  and dispersion 
 is generally used.  As a result,  questions naturally arise as to when a generic finite  slab can be regarded as an infinite substrate on which the theoretical calculations are based and how the novel out of thermal equilibrium effects depend on the  dielectric property.  
 In this paper, we try to answer these  questions in terms of the thermalization of a polarizable two-level atom in a thermal bath  near a planar dielectric slab out of thermal equilibrium.   We will show that for a  nonabsorbing and nondispersive dielectric  with a finite thickness NO out of thermal equilibrium  effects appear as far as the thermalization of the atom is concerned. So, to have non vanishing out of thermal equilibrium effects, one has to have a real dielectric substrate with an infinite thickness or a complex dielectric substrate. Since the infinitely thick substrate does not really exist, we give the condition when a dielectric with a tiny nonzero imaginary part in the relative permittivity with a finite thickness can be regarded as a  half-space dielectric. This puts on the solid foundation the experimental test using a finite dielectric substrate of theoretical predictions for novel effects from out of thermal equilibrium based upon a half-space dielectric.

\section{Open quantum system}
We examine in the framework of open quantum systems the thermalization of a two-level  atom near a dielectric substrate in a stationary configuration out of thermal equilibrium. We assume that  two stationary states  of  the atom are  represented by $|1\rangle$ and $|2\rangle$ respectively, and the energy spacing is $\hbar\omega_0$. A planar dielectric slab with thickness $d$ is placed in a thermal bath at temperature $T_0$  and its right  surface   coincides with the $z=0$ plane. The slab is assumed to be in local thermal equilibrium at  a  different temperature  $T_1$ (See Fig. 1).  The atom is at the position $z_A>0$ in the empty space. So, the whole system is out of thermal equilibrium but in a stationary regime and  the total Hamiltonian that governs the evolution of the system takes the  form 
\begin{eqnarray}\label{H1}
{H}={H}_A+{H}_B+{H}_{I}\;,
\end{eqnarray}
where ${H}_A=\sum_{m=1}^2 \omega_m |m\rangle\langle m|$ is  the Hamiltonian of the atom,  ${H}_B$ is the Hamiltonian describing the environment the atom is coupled to, and ${H}_{I}$ denotes the interaction between the atom and the environment, which takes the form ${H}_{I}= -{\vect{D}(t)}\cdot
{\vect{E}}(\vect{r}, t)$ in the multipolar coupling scheme. Here, ${\vect{D}(t)}$ is the electric dipole moment of the atom, and ${\vect{E}}(\vect{r}, t)$ is the electric field strength. 

In the interaction picture, the total density matrix $\rho_{tot}(t)$ of the system satisfies the von Neumann equation 
\bea{Neu}
\frac{\dif}{\dif t}\rho_{\text{tot}}(t)=-\frac{\mi}{\hbar} [H_I, \rho_{\text{tot}}(t)]\;,
\eea
with the initial state being described by $\rho_{tot}(0)=\rho(0)\otimes\rho_B$, where $\rho(0)$ is the initial density matrix of the atom and $\rho_B$ is that of the environment.  $H_I$ can be rewritten as \bea{HI} H_I=-\sum_{i, \omega} e^{-\mi\omega t} A_{i}(\omega) E_i(\mathbf{r}, t)\;,\eea
 where $i\in \{x, y, z\}$ and $A_{i}(\omega)=\sum_{\varepsilon'-\varepsilon=\omega} \Pi(\varepsilon)D_i \Pi(\varepsilon')$. $\Pi(\varepsilon)$ denotes the projection onto the eigenspace belonging to the eigenvalue $\varepsilon$ of $H_A$, which means that $A_i(\omega)$ are the eigenoperators of $H_A$.   
Tracing $\rho_{\text{tot}}(t)$ over the degrees of freedom associated with the environment, one can obtain the reduced density matrix $\rho(t)$ for the atom, namely,  $\rho(t)=\operatorname{Tr}_B\,[\rho_{\text{tot}}(t)]$, which, in the limit of weak coupling,  obeys the master equation
\bea{Mas}
\frac{\dif}{\dif t}\rho(t)=-\frac{\mi}{\hbar} [H_{LS}, \rho(t)]+\mathcal{D}(\rho(t))\;,
\eea
where 
\bea{HLS} H_{LS}=\hbar \sum_\omega\sum_{i,j} s_{ij}(\omega) A_i^\dag (\omega)A_j(\omega)\eea
is the so-called Lamb-shift Hamiltonian since it  produces  shifts of the atomic energy levels, and  
\bea{Drt}\mathcal{D}(\rho(t))= \sum_\omega\sum_{i,j} \gamma_{ij}(\omega) [A_j(\omega)\rho(t)A_i^\dag (\omega)-\frac{1}{2}\{A_i^\dag (\omega)A_j(\omega), \rho(t)\}]
\eea is the dissipator.  Here,   
\bea{gamma} \gamma_{ij}(\omega)=\frac{1}{\hbar^2} \int_{-\infty}^\infty \dif s e^{\mi\omega s}\langle E_i(\bm{r}, s)E_j(\bm{r},0)\rangle \eea is the Fourier transforms to the reservoir correlation function $\langle E_i(\bm{r}, s)E_j(\bm{r},0)\rangle$, while $s_{ij}$ is related to $\gamma_{ij}$ through 
\bea{sij} s_{ij}(\omega)=\frac{\mi}{2}\gamma_{ij}-\mi\tau_{ij}(\omega)\;, 
\eea 
where $\tau_{ij}(\omega)=\frac{1}{\hbar^2} \int_{0}^\infty \dif s e^{\mi\omega s}\langle E_i(\bm{r}, s)E_j(\bm{r},0)\rangle$ is the one-side Fourier transforms. With the help of  $\frac{1}{x\mp \mi \epsilon}=\mathrm{P}\frac{1}{x}\pm \mi \pi \delta(x)$, where $\mathrm{P}$ denotes the Cauchy principal value, $s_{ij}(\omega)$ can be re-expressed as 
\bea{sij11}s_{ij}(\omega)=-\frac{\mathrm{P}}{2\pi}\int_{-\infty}^{\infty} \frac{\gamma_{ij}(z)}{z-\omega}\dif z\;.\eea

For a two-level atom, the atomic dipole operator can be written as 
\bea{Dt} \vect{D}(t)=\vect{d}_{21}|2\rangle\langle 1|e^{-\mi\omega_0 t}+\vect{d}^*_{21}|1\rangle\langle 2|e^{\mi\omega_0 t}\;, \eea which implies that $\vect{A}(\omega)=\sum_{\omega} \vect{d}_{21}|2\rangle\langle 1|= \vect{A}^\dag(-\omega)$. Since the summation over $\omega$ just contains two terms: $\omega=\omega_0$ and $\omega=-\omega_0$, the master equation becomes 
\bea{Mast2}
\frac{\dif}{\dif t} \rho(t)&=&-\mi\bigg[ \sum_{n=1}^2 \omega_n |n\rangle \langle n| + S(\omega_0) |2\rangle \langle 2|+S(-\omega_0) |1\rangle \langle 1|, \rho(t)\bigg]\\ \nonumber
&+& \Gamma (\omega_0) \bigg( \rho_{22} |1\rangle \langle 1|-\frac{1}{2}\{|2\rangle \langle 2|, \rho(t)\}\bigg)+\Gamma (-\omega_0) \bigg( \rho_{11} |2\rangle \langle 2|-\frac{1}{2}\{|1\rangle \langle 1|, \rho(t)\}\bigg)\;,
\eea
where
\bea{2}
S(\omega_0)&\equiv& \sum_{i,j}s_{ij}(\omega_0)[\vect{d}_{21}]_i^*[\vect{d}_{21}]_j\;, \\ \nonumber
S(-\omega_0)&\equiv& \sum_{i,j}s_{ij}(-\omega_0)[\vect{d}_{21}]_i[\vect{d}_{21}]_j^*\;, \\  \label{3} 
\Gamma(\omega_0)&\equiv& \sum_{i,j}\gamma_{ij}(\omega_0)[\vect{d}_{21}]_i^*[\vect{d}_{21}]_j\;, \\  \nonumber
\Gamma(-\omega_0)&\equiv& \sum_{i,j}\gamma_{ij}(-\omega_0)[\vect{d}_{21}]_i[\vect{d}_{21}]_j^*\;.\eea
Here, $S(-\omega_0)$ and  $S(\omega_0)$ represent the atomic eigenvalue shifts of the ground state and the excited one, respectively, and the corresponding energy shifts are $\delta E_1=\hbar S(-\omega_0)$ and  $\delta E_2=\hbar S(\omega_0)$. As a result,   the Lamb shift (the relative energy shift) is given by $\Delta=\delta E_2-\delta E_1$.   $\Gamma(-\omega_0)$ and  $\Gamma(\omega_0)$  are the downward and upward transition rates respectively, which are related with the thermalization of an atom. Therefore, in the following only $\gamma_{ij}(\omega)$ is analyzed. 
From its definition,  we have 
\bea{Gamma1}
\gamma_{ij}(\omega)&=&\frac{1}{\hbar^2}\int_{-\infty}^{\infty}\dif s \int_0^\infty \dif \omega'\int_0^\infty \dif\omega''[e^{\mi(\omega-\omega'')s}\langle E_i(\vect{r},\omega'')E_j^\dag(\vect{r},\omega')\rangle\\ \nonumber  &&+e^{\mi(\omega+\omega'')s}\langle E_i^\dag(\vect{r},\omega'')E_j(\vect{r},\omega')\rangle]\\ \nonumber
&=& \frac{2\pi}{\hbar^2} \int_0^\infty \dif \omega'\begin{cases}\langle E_i(\vect{r},\omega)E_j^\dag(\vect{r},\omega')\rangle & \omega>0\\
\langle E_i^\dag(\vect{r},-\omega)E_j(\vect{r},\omega')\rangle & \omega<0\end{cases}\;,
\eea
where  $\langle E_i(\vect{r},\omega)E_j(\vect{r},\omega')\rangle=\langle E_i^\dag(\vect{r},\omega)E_j^\dag(\vect{r},\omega')\rangle=0$ and $\int_{-\infty}^\infty \dif s\exp(-\mi\epsilon s)=2\pi\delta(\epsilon)$ have been used.

For a nonmagnetic medium,  the electric field operator can be expressed as 
\begin{eqnarray}
\label{eq5}
{\vect{E}}(\vect{r}, \omega)
&\!=&\! 
 \mi\,\frac{\omega^2}{c^2}
 \sqrt{\frac{\hbar}{\pi\epsilon_0} }\int\dif^3r'\,
 \sqrt{ \operatorname{Im}\epsilon(\vect{r}',\omega)}\,
 \ten{G}(\vect{r},\vect{r}',\omega) \sprod {\vect{f}}(\vect{r}',\omega)\;.
\end{eqnarray} 
Here, $\epsilon_0$ and $\epsilon$ are the vacuum and relative permittivity, respectively,  $\ten{G}$ is  the classical Green's tensor, which  satisfies  an useful integral relation \begin{eqnarray}
\label{eq14}
\int\dif^3 s\,
 \operatorname{Im}\epsilon(\vect{s},\omega)\, \ten{G}(\vect{r},\vect{s},\omega)\!\cdot\!
 \ten{G}^{\ast\trans}(\vect{r}',\vect{s},\omega)
=\frac{c^2}{\omega^2}\operatorname{Im}
 \ten{G}(\vect{r},\vect{r}',\omega)\;,
\end{eqnarray} 
and  ${\vect{f}}(\vect{r},\omega)$ and
${\vect{f}}^\dagger(\vect{r},\omega)$  are the  annihilation and creation
 operators of  the elementary electric excitations, respectively. They obey the bosonic commutation relations
$ [{\vect{f}} (\vect{r},\omega),
{\vect{f}} ^\dagger(\vect{r}',\omega')] =\bm{\delta} 
(\vect r-\vect r') \delta(\omega-\omega') $
and
$ [{\vect{f}}(\vect{r},\omega), {\vect{f}}
(\vect{r}',\omega')] =[{\vect{f}}^\dagger(\vect{r},
\omega), {\vect{f}}^\dagger(\vect{r}',\omega')] 
=\ten{0},$
where $\ten{0}$ represents a zero matrix.  For the thermal state describing the system in a stationary configuration out of thermal equilibrium we are considering, one has 
\begin{equation}
\label{eq2c1}
\langle\{\beta_i\}| {\vect{f}}(\vect{r},\omega) {\vect{f}}^\dagger (\vect{r}',\omega')|\{\beta_i\}\rangle
=[1+N(\omega, \beta_i)] \delta(\vect{r}-\vect{r}')\delta(\omega-\omega')\;,
\end{equation}
\begin{equation}
\label{eq2c2}
\langle\{\beta_i\}| {\vect{f}}^\dagger(\vect{r},\omega) {\vect{f}} (\vect{r}',\omega')|\{\beta_i\}\rangle
=N(\omega, \beta_i) \delta(\vect{r}-\vect{r}') \delta(\omega-\omega')\;,
\end{equation}
where $\beta_i=\hbar c/kT_i$ with $i=0$ or $1$,   and
$N(\omega, \beta_i)=\frac{1}{e^{\beta_i \omega/c}-1}$.
Substituting Eq.~(\ref{eq5}) into Eq.~(\ref{Gamma1}) and considering the relations given in Eqs.~(\ref{eq2c1}, \ref{eq2c2}),  we have 
\bea{Gamma2}
\gamma_{ij}(\omega)=\frac{2\mu_0 \omega^4}{\hbar c^2} \int \dif^3 {r}'[1+N(\omega, \beta)]   \operatorname{Im}\epsilon(\vect{r}',\omega) {G}_{ ik} (\vect{r}, \vect{r}', \omega) {G}^*_{jk} (\vect{r}, \vect{r}',  \omega) \;,
\eea
and
\bea{Gamma3}
\gamma_{ij}(-\omega)=\frac{2\mu_0 \omega^4}{\hbar c^2} \int \dif^3 {r}'\,N(\omega, \beta)   \operatorname{Im}\epsilon(\vect{r}',\omega) {G}_{ik} (\vect{r}, \vect{r}', \omega) {G}^*_{jk} (\vect{r}, \vect{r}',  \omega) \;.
\eea
Here $\mu_0$ is the vacuum permeability and $\mu_0 \epsilon_0=1/c^2$ is used. 
\begin{figure}[t]
\scalebox{0.4}{\includegraphics{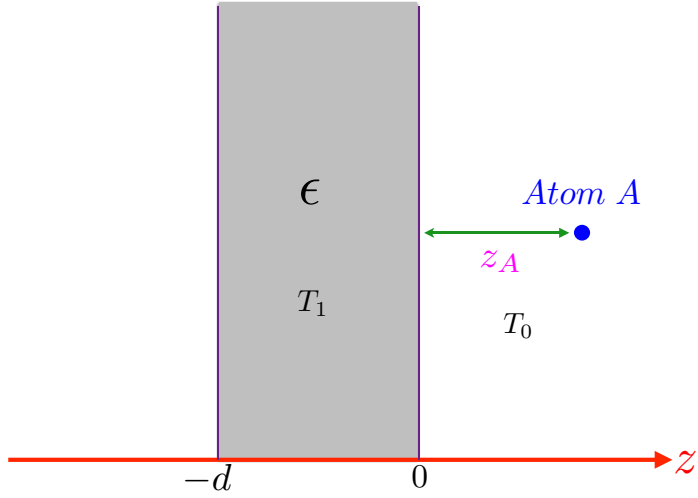}}
\caption{(color online). Scheme of the system considered.}
\label{Fig1}\end{figure}
For an atom near a    dielectric slab described in Fig.~\ref{Fig1}, Eqs.~(\ref{Gamma2}, \ref{Gamma3}) become
\bea{Gamma4}
\gamma_{ij}(\omega)&=& \frac{2  \mu_0 \omega^2}{\hbar}    [1+N(\omega, \beta_0)] \operatorname{Im} G_{ij}(\vect{r}_A, \vect{r}_A, \omega) \\ \nonumber 
&&+ \frac{2 \pi }{\hbar}      [N(\omega, \beta_1)-N(\omega, \beta_0)]\, g_{ij} (\vect{r}_A, \vect{r}_A, \omega )\;, 
\eea
and
\bea{Gamma5}
\gamma_{ij}(-\omega)&=& \frac{2\mu_0  \omega^2}{\hbar}    N(\omega, \beta_0) \operatorname{Im} G_{ij}(\vect{r}_A, \vect{r}_A, \omega)  \\ \nonumber  && +\frac{2\pi }{\hbar}  \,   [N(\omega, \beta_1)-N(\omega, \beta_0)]\, g_{ij} (\vect{r}_A, \vect{r}_A, \omega ) \;,
\eea
where Eq.~(\ref{eq14}) has been used,   and 
\bea{gij} g_{ij}(\vect{r}, \vect{r}, \omega )\equiv \frac{\mu_0\omega^4 }{\pi c^2}\int \dif^2 \vect{r}'_{\|}  \int^{0}_{-d}  \dif z'   \operatorname{Im}\epsilon \, {G}_{ik} (\vect{r}, \vect{r}', \omega) {G}^*_{jk} (\vect{r}, \vect{r}',  \omega)\;,
\eea
with $\vect{r}'_{\|} =\{ x', y'\}$.  In the right hand side of Eqs.~(\ref{Gamma4}, \ref{Gamma5}), the first term gives  the contributions of the zero-point fluctuations and  the thermal fluctuations in thermal equilibrium 
at a temperature $T_0$,  while the second term arises from the out of thermal equilibrium nature of the system. For the  system we are considering, only the  diagonal elements of  $\operatorname{Im} G_{ij}(\vect{r}, \vect{r}, \omega) $ and $g_{ij} (\vect{r}, \vect{r}, \omega )$   are non-vanishing.

\section{Thermalization}
Using Eqs.~(\ref{Gamma4}, \ref{Gamma5}), we can show  that the transition rates $\Gamma(\omega_0)$ and $\Gamma(-\omega_0)$ can be re-expressed as
\begin{equation}\label{GammaN}\begin{split}\begin{pmatrix}\Gamma(\omega_{0})\\\Gamma(-\omega_{0})\end{pmatrix}&=\alpha(\omega_0) \Gamma_0(\omega_{0})  \begin{pmatrix}1+N_\text{eff} (\omega_0)\\N_\text{eff}(\omega_0)\end{pmatrix} ,\end{split}\end{equation}
where $\Gamma_0(\omega_{0})=\frac{\omega_{0}^3|\mathbf{d}_{12}|^2}{3\pi\epsilon_0\hbar c^3}$ is the vacuum spontaneous-emission rate related to the transition between the ground and exited states,  \bea{ome0} \alpha(\omega_0)= \frac{6\pi c}{ \omega_0}  \sum_{i,j}  \frac{[\vect{d}_{21}]_i[\vect{d}_{21}]_j^*}{\modul{\vect{d}_{21}}^2 }\text{Im} G_{ij}(\vect{r}, \vect{r}, \omega_0) \;, \eea
 and 
\bea{nf}
  N_\text{eff}(\omega_0) &=&  N(\omega_0, \beta_0)  +\frac{6 \pi^2 c} {\mu_0 \omega_0^3  \alpha(\omega_0) }    [N(\omega_0, \beta_1)-N(\omega_0, \beta_0)]  \sum_{i,j} \frac{[\vect{d}_{21}]_i[\vect{d}_{21}]_j^*}{\modul{\vect{d}_{21}}^2 }g_{ij} (\vect{r}_A, \vect{r}_A, \omega_0) \nonumber  \\ 
&=& N(\omega_0, \beta_0)  +\frac{2 \pi^2  c} {\mu_0 \omega_0^3  \alpha(\omega_0) }  \,   [N(\omega_0, \beta_1)-N(\omega_0, \beta_0)] \operatorname g (\vect{r}_A, \vect{r}_A, \omega_0)\;.
\eea
Here, the last line holds for an isotropically polarizable atom, and $g =g_{xx}+ g_{yy}+ g_{zz}$. So, $N_\text{eff}(\omega_0)$ depends on the temperature $T_i$ ($i=0, 1$) and the dielectric property of slab encoded in function $g (\vect{r}, \vect{r}, \omega )$.  As discussed in~\cite{Bellomo, Hu13}, after evolving for a sufficiently long period of time, the atom will be  thermalized to a steady state with an effective temperature 
\bea{T}
T_\text{eff}=\frac{\hbar \omega_0}{k} [\ln (1+N_\text{eff}^{-1}(\omega_0))]^{-1}\;.
\eea
It is easy to see that if the  substrate is in thermal equilibrium with the thermal radiation in the empty space where the atom is located,  then $T_\text{eff} $ reduces to $T_0$ as expected. 

In order to analyze in detail the thermalization temperature of the atom, we first need to examine the behavior of $\operatorname g (\vect{r}, \vect{r}, \omega )$, which depends on the Green's function     $\ten{G}(\omega, \vect{r}, \vect{r'})$, where  $\vect{r}$ indicates the position of the atom  and thus it is restricted to the empty right half-space, while $\vect{r}'$ is in the slab.  For the system considered, the Green's function  can be expanded as \bea{G10}
\ten{G}(\omega, \vect{r},\vect{r}')=\int d^2 \vect{k}\; e^{\mi\vect{k} \cdot(\vect{r}_{\|}-\vect{r}'_{\|}) } \ten{G}(\vect{k}, \omega, z, z')
\eea
where $\vect{k}=(k_x, k_y)$.  Since $z$ and $z'$ are  in different regions,  from Refs.~\cite{Tomas1995, Chew} we have that
\bea{G30}
\ten{G}(\vect{k}, \omega, z, z')&=&\frac{\mi}{8\pi^2 b_0(k)}\sum_{\sigma=s,p} \xi^\sigma \frac{t^\sigma (k) e^{\mi b_0(k)z }}{D^{\sigma}(k)}   \cdot \nonumber \\
&& \hat{\vect{e}}_{\sigma_{0}}^+(\vect{k}) \big [
 \hat{\vect{e}}_{\sigma_1}^- (-\vect{k}) e^{-\mi  b_1 (k) z' }+r^{\sigma}_{-} (k)\hat{\vect{e}}_{\sigma_1}^+ (-\vect{k}) e^{\mi b_1 (k) (z'+2d )} \big]  \;,  
\eea
where  $\xi^p=1$, $\xi^s=-1$,  $b_0(k)= \sqrt{k_{0}^{2}-k^2}$,  $b_1(k)= \sqrt{k_1^{2} -k^2}$, $k=\modul{\vect{k}}$,  $k_{0}= \frac{\omega}{c}$, $k_{1}= \sqrt{\epsilon}\frac{\omega}{c}$, and  \bea{D}D^{\sigma}(k)=1-  r^{\sigma}_{-} r^{\sigma}_{+}  (k) e^{2\mi b_1(k)  d}\,, \nonumber
\eea 
with   $r^{\sigma}_{+} (k)$ and $r^{\sigma}_{-} (k)$ being the  reflection coefficients  at the right and left boundaries of the slab,  which have the forms:  
\bea {rs} r^s_{\pm}  (k) = \frac{b_1 (k) -b_{0} (k)}{b_1 (k)+b_{0} (k)}\;, \quad  r^p_{\pm} (k) = \frac{b_1 (k) -\epsilon b_{0} (k)}{b_1 (k) + \epsilon b_{0} (k)}\;. \eea
$t^\sigma (k)=\sqrt{\frac{1}{\epsilon}}(1-r^{\sigma}_{+}(k))$ is the transmission coefficient between the empty space  and the slab. In addition, we define 
\bea{eps} \hat{\vect{e}}_{p_{i}}^\pm (\vect{k})  ={\frac{1}{k_{i}}}(\mp b_i \hat{\vect{k}}+k \hat{\vect{z}})\;,\quad  \hat{\vect{e}}_{s_{i}}^\pm  (\vect{k}) =\hat{\vect{k} } \times \hat{ \vect{z} }\;.  \eea

\begin{figure}[t]
\scalebox{0.5}{\includegraphics{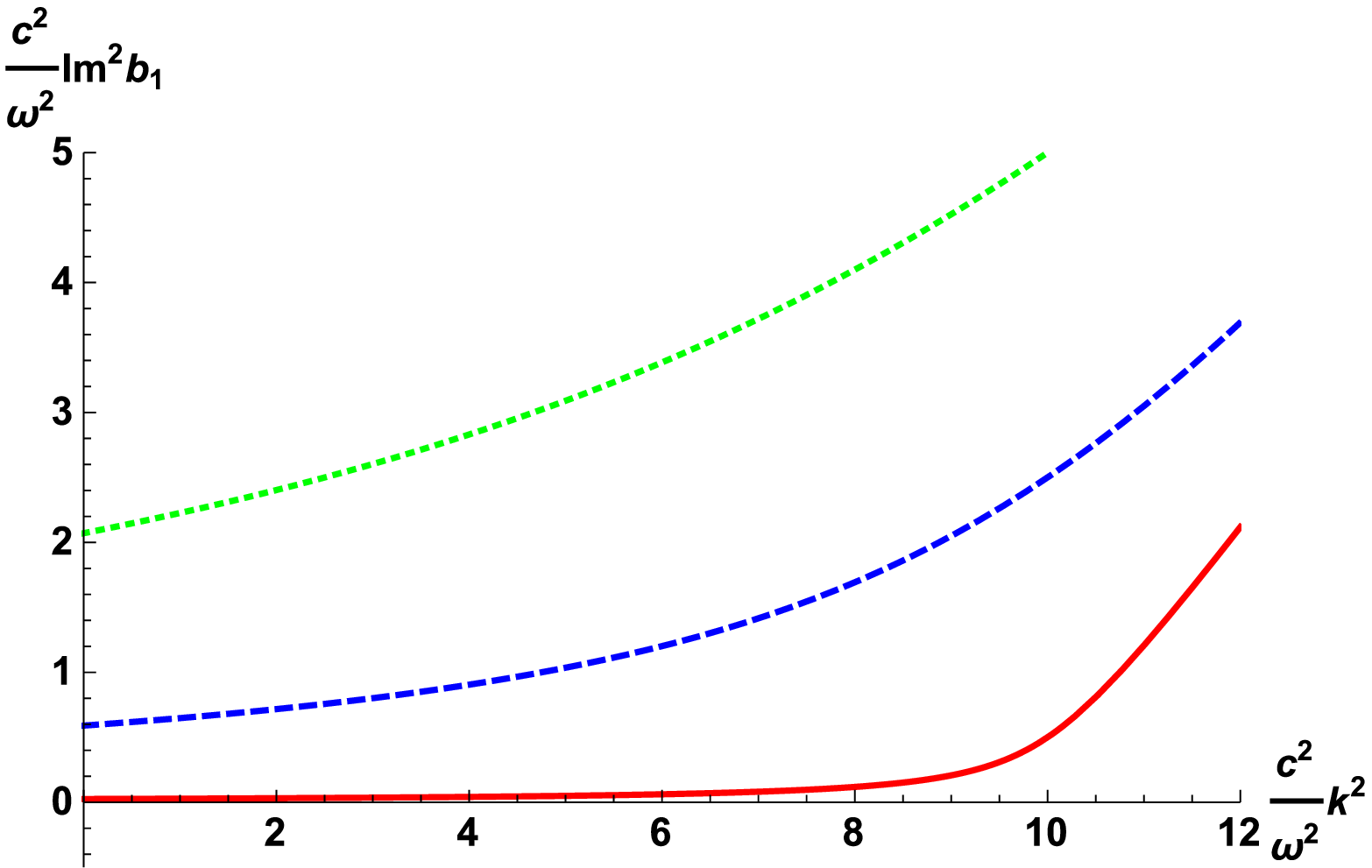}}\scalebox{0.5}{\includegraphics{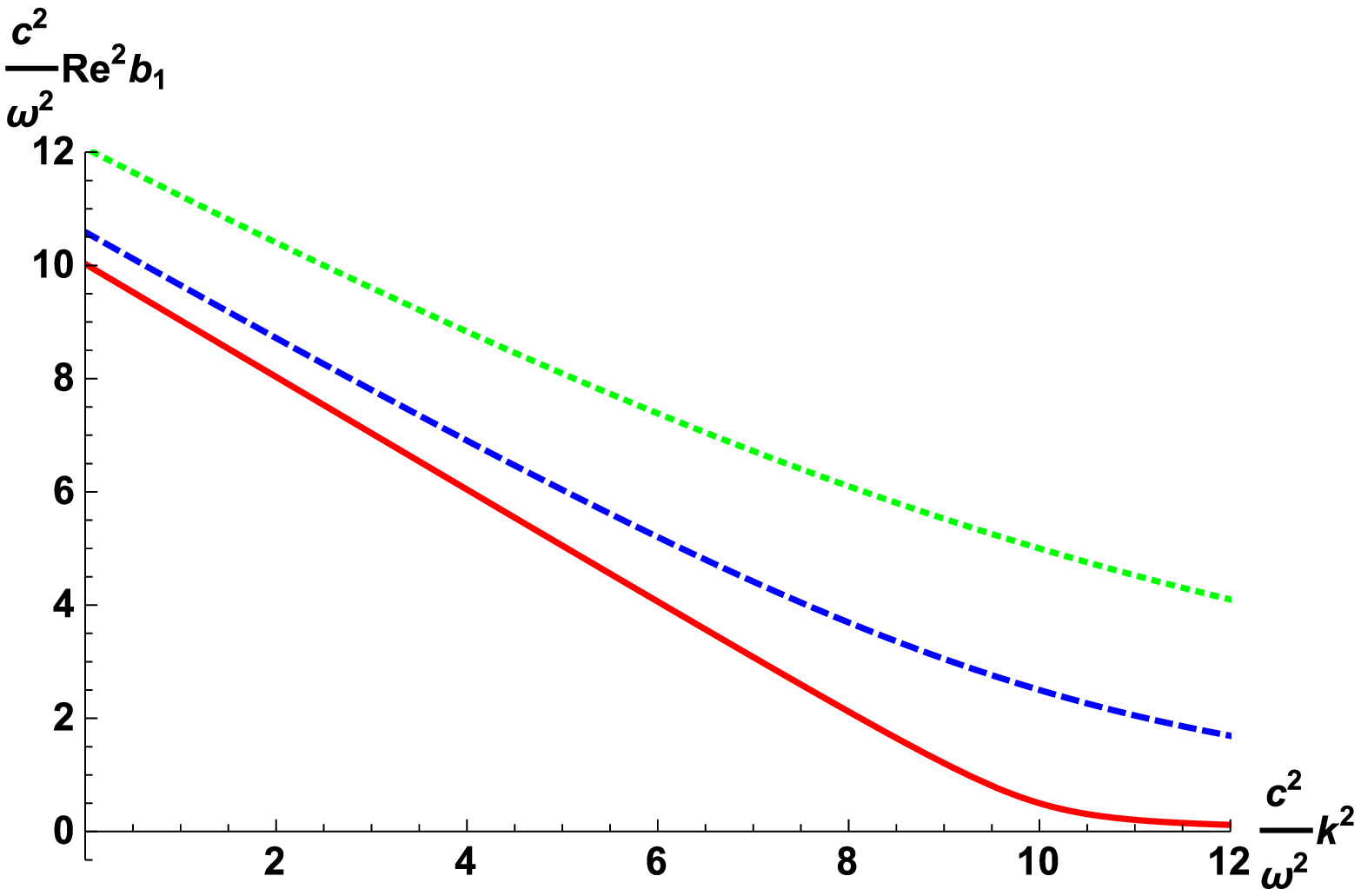}}
\caption{(color online). The evolutionary curves of $\frac{c^2}{w^2}\operatorname{Im}^2 b_1$ (left panel) and $\frac{c^2}{w^2}\operatorname{Re}^2 b_1$ (right panel) with respect to  $\frac{c^2}{w^2} k^2$ with $ \operatorname{Re}\epsilon=10 $. The red solid, blue dashed and green dotted lines correspond to $ \operatorname{Im}\epsilon=1 $, $5$ and $10$, respectively. }
\label{Fig2}\end{figure}

Substituting Eqs.~(\ref{G10}, \ref{G30}) into Eq.~(\ref{gij}), for a system described in Fig.~\ref{Fig1} and an isotropically polarizable atom,  one has $\operatorname  g(\vect{r}, \vect{r}, \omega )=g(z, z, \omega )$  with 
\bea{gii2}
g(z, z, \omega )&=&\frac{\mu_0 \omega^2}{8\pi^2} \int_0^\infty \frac{ k\dif k}{\modul{b_0 (k) }^2} e^{-2 \operatorname{Im} b_0(k)  z  } \bigg \{ \operatorname{Re} b_1 (k) [A_+(k) +{A}(k)] (1-  e^{-2\operatorname{Im} b_1(k)   d } )  \nonumber \\
&&+ e^{-2\operatorname{Im} b_1(k)   d }\bigg(\operatorname{Re} b_1 (k) [A_+(k) \modul{r^p_{-} (k)}^2+{A}(k)\modul{r^{s}_- (k)}^2]   (1- e^{- 2\operatorname{Im} b_1 (k)  d }) \nonumber \\
&&+  2\operatorname{Im}b_1 (k) [A_- (k) \operatorname{Re}r^{p}_{-}(k)+ {A}(k)\operatorname{Re}r^{s}_- (k) ]  \sin(2\operatorname{Re}b_1 (k)  d )
  \\ \nonumber
&& +2\operatorname{Im}b_1 (k) [A_-(k) \operatorname{Im}r^{p}_{-}(k)+ {A}(k) \operatorname{Im}r^{s}_{-}(k) ]  [\cos(2\operatorname{Re}b_1(k)  d)-1] \bigg)\bigg\}\;.
\eea
where \bea{Ak} A_{\pm} (k)=\modul{\frac{t^p (k)}{D^{p}(k)}}^2  \frac{[k^2\pm \modul{b_1(k)}^2][k^2+\modul{b_0(k)}^2]}{\modul{k_0 k_1}^2}\;,\quad {A}(k)=\modul{\frac{t^s(k)}{D^{s}(k)}}^2 \;.\eea   This expression shows that $\operatorname{Im} b_0 (k)$ must be nonzero, otherwise $g(z, z, \omega )$ will become a constant independent of $z$.  From the definition of  $b_0({k})$ and $b_1({k})$, we obtain that  
\bea{b0} 2 \operatorname{Im}^2 b_0(k)=-\big(\frac{\omega^2}{c^2}-k^2 \big)+ \modul{\frac{\omega^2}{c^2}-k^2}
\eea and
\bea {bl}
\operatorname{Im}^2 b_1(k)&=&\frac{1}{2}\bigg [-\bigg(\frac{\omega^2}{c^2} \operatorname{Re} \epsilon-k^2\bigg)+ \sqrt{\frac{\omega^4}{c^4} \operatorname{Im}^2\epsilon + \bigg(\frac{\omega^2}{c^2} \operatorname{Re} \epsilon-k^2\bigg)^2 }\,\bigg ]\;,\\ 
 \operatorname{Re}^2 b_1 (k)&=& \frac{1}{2}\bigg [\bigg(\frac{\omega^2}{c^2} \operatorname{Re} \epsilon-k^2\bigg)+ \sqrt{\frac{\omega^4}{c^4} \operatorname{Im}^2\epsilon + \bigg(\frac{\omega^2}{c^2} \operatorname{Re} \epsilon-k^2\bigg)^2 }\,\bigg ]  \label{bl2} \;.
\eea
A nonzero $\operatorname{Im} b_0(k)$   means  that $k^2>\frac{\omega^2}{c^2}$ and thus only the $k>\frac{\omega}{c}$ interval in $k$ integration from $0$ to $\infty$ contributes. Let us note that  $\operatorname{Im}^2 b_1(k)$ is an increasing function  of  $k^2$, while $\operatorname{Re}^2 b_1(k)$ is a decreasing one, as is shown graphically in Fig.~\ref{Fig2}.

If the slab consists of real dielectrics, i.e.,  $\operatorname{Im} \epsilon=0$, Eqs.~(\ref{bl}, \ref{bl2}) tell us  that  $\operatorname{Im}b_1(k)= 0$, if $\operatorname{Re}b_1 (k)\neq 0$,  and vice versa. As a result,  it is easy to see that  $g (z, z, \omega )=0$.
Only when the slab thickness is infinite, that is,  $d \rightarrow \infty$ and $r^{\sigma}_{-}=0$,   is $g(z, z, \omega )$ nonzero     
and it then becomes
\bea{gii3}
g(z, z, \omega )&=&\frac{\mu_0 \omega^2}{8\pi^2} \int_0^\infty \frac{ k\dif k}{\modul{b_0 (k) }^2} e^{-2 \operatorname{Im} b_0(k)  z  }  \operatorname{Re} b_1 (k) [\bar{A}_+(k) +\bar{A}(k)] \eea
where \bea{Abk} \bar{A}_{+} (k)=\modul{t^p (k)}^2 \frac{[k^2+ \modul{b_1(k)}^2][k^2+\modul{b_0(k)}^2]}{\modul{k_0 k_1}^2}\;,   \quad \bar{A}(k)=\modul{ t^s(k)}^2 \;.\eea  This demonstrates that there is NO out of thermal equilibrium effect for any real dielectric substrate of finite thickness even when  the 
substrate is held at a different local temperature.  In other words, an infinite thickness is the only way to have an out thermal equilibrium effect for a 
real dielectric substrate.

Another way to have a nonzero out of thermal equilibrium effect is that the slab consists of the dispersive and absorbing dielectric ($\operatorname{Im}\epsilon\neq 0$).  This  is similar to what happens to the decay rate of the exited state of an atom in front of a dielectric plate, which is proportional to the imaginary part of the permittivity and  also equals to zero when $\operatorname{Im}\epsilon= 0$~\cite{DF}.  From the Eq.~(\ref{gii2}), one can see that, if $2\operatorname{Im} b_1 (k) d>1$,   the  terms depending on $d$  can be neglected since they are exponentially suppressed  as compared to the other term, which then gives the dominant  contribution. In this case,  the result of 
 the integral becomes effectively independent of $d$ and approximates to that  in the case of a half-space dielectric substrate, which has the same form as  that given in Eq.~(\ref{gii3}). Since $\operatorname{Im}^2 b_1 (k)$ is an increasing function  of  $k^2$ and $k^2>\frac{\omega^2}{c^2}$ is required,  the  minimum value of  $\operatorname{Im} b_{1} (k) $ is achieved at $k^2=\frac{\omega^2}{c^2}$ 
\bea{MIm}
{\text{Min}}\{\operatorname{Im} b_1 (k)\}=\frac{1}{\sqrt{2}}\frac{\omega}{c}\bigg [-( \operatorname{Re} \epsilon-1)+ \sqrt{ \operatorname{Im}^2 \epsilon+ ( \operatorname{Re} \epsilon-1)^2 }\bigg ]^{1/2} \;.
\eea
So, the condition for the thermalization of  an two-level atom with a typical transition $\omega_0$ near a dielectric slab of finite thickness $d$ out of thermal equilibrium to behave like that near an infinitely thick half-space dielectric  substrate  is
\bea{Cond}
 \frac{\sqrt{2}  d}{\lambda_0}  \bigg [-( \operatorname{Re} \epsilon-1)+ \sqrt{ \operatorname{Im}^2 \epsilon+ ( \operatorname{Re} \epsilon-1)^2 }\bigg ]^{1/2} > 1\;,
\eea
where  $\lambda_0=\frac{c}{\omega_0}$ is the transition wavelength of the atom.  Since the dielectrics with a very small but nonzero $\operatorname{Im}  \epsilon$, such as fused silica and sapphire,  are  used  in the experiment to observe the novel feature for the CP force out of thermal equilibrium~\cite{Obrecht}, we expand the condition in the limit of $\operatorname{Im} \epsilon\sim 0$ and obtain
\bea{M}\frac{\operatorname{Im} \epsilon}{\sqrt{\operatorname{Re}\epsilon-1}} \frac{d}{\lambda_0} > 1\;.\eea
Obviously, for a given atom we can always find a finite $ d$ so that the condition is satisfied as long as $ \operatorname{Im} \epsilon$ is not vanishing no matter how small it is.  Since mathematically infinite thick slab does not exist,  the above relation can serve as a guide for an experimental verification of the effects that arise from out of thermal equilibrium fluctuations, and  makes it justified to test experimentally the novel property  theoretically found from a half-space  dielectric out of thermal equilibrium using a dielectric substrate of  finite thickness with a tiny imaginary part in the relative permittivity.   

Now a few comments are in order. First, our results  can be generalized to a system of a multilayer dielectric body with each layer of a different permittivity in local thermal equilibrium at a different temperature.  For the multilayered substrate consisting of only real dielectric, if the outermost left layer is a perfect mirror or empty space, the system has no out of thermal equilibrium effect at least as far as the thermalization of the atom is concerned. If the outermost left layer is a half-space substrate  at a certain temperature,  only this temperature and that  of the thermal bath in the right half  empty space affect  the thermalization of the atom.  Second, although, our calculations are performed under the assumption of an isotropically 
polarizable atom,  our conclusions also hold  for an anisotropically polarizable atom since the only difference for such a case is that the definitions of ${A}_{\pm}(k)$ and ${A}(k)$ in Eq.~(\ref{gii2}) are different. Finally, here we only investigate the thermalization of an atom in front of a slab. The out of thermal equilibrium Casimir-Polder force, especially its  explicit dependence   on the thickness $d$, the distance $z_{A}$ and the temperatures $T_{i}$ in different limits like what was discussed in~\cite{DF} for the thermal Casimir-Polder force is an interesting topic which is currently under investigation.

\section{Conclusion}

In conclusion, we have studied the thermalization of a two-level atom near a planar dielectric substrate in a stationary environment out of thermal equilibrium in which the atom is located in an empty space filled with a thermal bath at a temperature different  from the local thermal equilibrium temperature of the substrate. We demonstrate that when the planar dielectric substrate is a real dielectric of finite thickness, no out of thermal equilibrium effects appear as far as the thermalization of the atom is concerned. That is to say, the atom thermalizes as if the substrate is in thermal equilibrium with 
the thermal bath in the empty space where the atom is located.  We also show that a planar  dispersive and absorbing dielectric substrate with a finite thickness and a tiny imaginary part in the relative permittivity, in its influence on the thermalization of the atom, behaves like a half-space  dielectric  under certain condition, and we concretely derive this condition  in our paper, which  can serve as a guide for an experimental verification, using a dielectric substrate of a finite thickness, of the effects that arise from out of thermal equilibrium fluctuations with a half-space (infinite thickness)  dielectric.

\acknowledgments  We acknowledge Wenting Zhou and Jiawei Hu for useful discussions.  This work was supported by the National Natural Science Foundation of China under Grants No. 11175093, No. 11222545, No. 11435006, and No. 11375092;  the  Specialized Research Fund for the Doctoral Program of Higher Education under Grant No. 20124306110001; 
 and the K.C. Wong  Magna Fund of Ningbo University.
 


\begin{thebibliography}{99}
\bibitem{Antezza}M. Antezza, L. P. Pitaevskii, and S. Stringari, Phys. Rev. Lett.
{\bf 95}, 113202 (2005);  M. Antezza, J. Phys. A {\bf 39}, 6117 (2006).
\bibitem{Casimir} H. B. G. Casimir and D. Polder, Phys. Rev. {\bf 73}, 360 (1948).
\bibitem{Lifshitz}E. M. Lifshitz, Zh. Eksp. Teor. Fiz. {\bf 29}, 94 (1955); Sov. Phys. JETP {\bf 2}, 73 (1956).
\bibitem{Zhou014} W. Zhou and H. Yu, Phys. Rev. A {\bf 90}, 032501 (2014). 
\bibitem{Wu14}P. Wu and H. Yu, Phys. Rev. A {\bf 90}, 032502 (2014).
\bibitem{Obrecht} J. M. Obrecht, R. J. Wild, M. Antezza, L. P. Pitaevskii, S.
Stringari, and E. A. Cornell, Phys. Rev. Lett. {\bf 98}, 063201 (2007). 
\bibitem{Bellomo}B. Bellomo, R. Messina, D. Felbacq, and M. Antezza, Phys. Rev. A {\bf 87}, 012101 (2013); EPL {\bf 100}, 20006 (2012).
\bibitem{Hu13}J. Hu, W. Zhou, and H. Yu, Phys. Rev. D {\bf 88}, 085035 (2013).
\bibitem{Bellomo13a} B. Bellomo and M. Antezza, EPL {\bf104}, 10006 (2013).
\bibitem{Bellomo13b} B. Bellomo and M. Antezza, New J. Phys. {\bf15}, 113052 (2013); Phys. Rev. A {\bf 91}, 042124 (2015).
\bibitem{Joulain} K. Joulain, et al., Surf. Sci. Rep. {\bf57}, 59 (2005).
\bibitem{Messina12}R. Messina, M. Antezza, and P. Ben-Abdallah, Phys. Rev. Lett. {\bf109}, 244302 (2012).
\bibitem{Leggio}B. Leggio, B. Bellomo, and M. Antezza, Phys. Rev. A {\bf91}, 012117 (2015)
\bibitem{Pitaevskii}L. Pitaevskii, J. Phys. A: Math. Gen. {\bf 39}, 6665 (2006). 
\bibitem{Antezza08}M. Antezza, L. P. Pitaevskii, S. Stringari, and V. B. Svetovoy, Phys. Rev. A {\bf 77}, 022901 (2008).
\bibitem{Buhmann}S. Y.  Buhmann and S. Scheel, Phys. Rev. Lett. {\bf100}, 253201 (2008); Phys. Scr. {\bf T135}, 014013 (2009).
\bibitem{Sherkunov}Y. Sherkunov, Phys. Rev. A {\bf 79},  032101 (2009).
\bibitem{Bimonte}G. Bimonte, Phys. Rev. A {\bf 80},  042102 (2009); Phys. Rev. A {\bf 92}, 032116 (2015).
\bibitem{Rodriguez}  J. J. Rodriguez and A. Salam, Phys. Rev. A {\bf 82}, 062522 (2010).
\bibitem{Druzhinina}V. Druzhinina, M. Mudrich, F. Arnecke, J. Madronero, and  A. Buchleitner, Phys. Rev. A {\bf 82}, 032714 (2010).
\bibitem{Behunin}R. O. Behunin and B. L. Hu, Phys. Rev. A {\bf 82}, 022507 (2010); Phys. Rev. A {\bf 84}, 012902 (2011). 
\bibitem{Messina} R. Messina and M. Antezza, EPL {\bf 95}, 61002 (2011); Phys. Rev. A {\bf 84}, 042102 (2011). 
\bibitem{Kruger}M. Kruger, T. Emig, G. Bimonte, and M. Kardar, EPL {\bf 95}, 21002 (2011).
\bibitem{Kruger11}M. Kruger, T. Emig, and M. Kardar, Phys. Rev. Lett. {\bf 106}, 210404 (2011).
\bibitem{Zhou15} W. Zhou and H. Yu, Phys. Rev. A {\bf 91}, 052502 (2015).
\bibitem{Noto}  A. Noto, R. Messina, B. Guizal, and M. Antezza, Phys. Rev. A {\bf 90}, 022120 (2014).
\bibitem{Antezza06} M. Antezza, L. P. Pitaevskii, S. Stringari, and V. B. Svetovoy,
Phys. Rev. Lett. {\bf 97}, 223203 (2006).
\bibitem{Tomas1995}M. S. Tomas, Phys. Rev. A {\bf 51}, 2545 (1995).
\bibitem{Chew}W. C. Chew, {\it Waves and fields in inhomogeneous media},  IEEE Press, 1995.
\bibitem{DF}S. Y. Buhmann, {\it Dispersion Forces II},  Springer Press, 2012.

\end{thebibliography}
\end{document}